\def \SAIT #1 #2 {{\em Mem.\ Soc.\ Astron.\ It.\/} {\bf #1}, #2}
\def \MESS #1 #2 {{\em The Messenger\/} {\bf #1}, #2}
\def \ASTRNACH #1 #2 {{\em Astron. Nach.\/} {\bf #1}, #2}
\def \AAP #1 #2 {{\em Astron. Astrophys.\/} {\bf #1}, #2}
\def \AAL #1 #2 {{\em Astron. Astrophys. Lett.\/} {\bf #1}, L#2}
\def \AAR #1 #2 {{\em Astron. Astrophys. Rev.\/} {\bf #1}, #2}
\def \AAS #1 #2 {{\em Astron. Astrophys. Suppl. Ser.\/} {\bf #1}, #2}
\def \AJ #1 #2 {{\em Astron. J.\/} {\bf #1}, #2}
\def \ANNREV #1 #2 {{\em Ann. Rev. Astron. Astrophys.\/} {\bf #1}, #2}
\def \APJ #1 #2 {{\em Astrophys. J.\/} {\bf #1}, #2}
\def \APJL #1 #2 {{\em Astrophys. J. Lett.\/} {\bf #1}, L#2}
\def \APJS #1 #2 {{\em Astrophys. J. Suppl.\/} {\bf #1}, #2}
\def \APSS #1 #2 {{\em Astrophys. Space Sci.\/} {\bf #1}, #2}
\def \ASR #1 #2 {{\em Adv. Space Res.\/} {\bf #1}, #2}
\def \BAIC #1 #2 {{\em Bull. Astron. Inst. Czechosl.\/} {\bf #1}, #2}
\def \JSQRT #1 #2 {{\em J. Quant. Spectrosc. Radiat. Transfer\/} {\bf #1}, #2}
\def \MN #1 #2 {{\em Mon. Not. R. Astr. Soc.\/} {\bf #1}, #2}
\def \MEM #1 #2 {{\em Mem. R. Astr. Soc.\/} {\bf #1}, #2}
\def \PLR #1 #2 {{\em Phys. Lett. Rev.\/} {\bf #1}, #2}
\def \PASJ #1 #2 {{\em Publ. Astron. Soc. Japan\/} {\bf #1}, #2}
\def \PASP #1 #2 {{\em Publ. Astr. Soc. Pacific\/} {\bf #1}, #2}
\def \NAT #1 #2 {{\em Nature\/} {\bf #1}, #2}
\documentstyle[12pt]{article}
\input epsf.sty
\newcommand{\be}{\begin{equation}}
\newcommand{\ee}{\end{equation}}

 1
\font\elevenrm=cmr10 scaled\magstep 1
 1

\def\refe{\hang\noindent}

\begin{document}
\vspace*{1.8cm}
  \centerline{\bf SIGNATURES OF RESONANT ABSORPTION} 
  \centerline{\bf IN X-RAY SPECTRA OF TYPE 1 AGN}
\vspace{1cm}
  \centerline{F. Nicastro$^{1,2,3}$, F. Fiore$^{1,3}$, G. Matt$^4$}
\vspace{1.4cm}
  \centerline{$^1$Harvard-Smithsonian Astrophysical Observatory}
  \centerline{\elevenrm 60 Garden Street, Cambridge MA 02138, USA}
  \centerline{$^2$Istituto di Astrofisica Spaziale -- CNR}
  \centerline{\elevenrm Via del Fosso del Cavaliere, 00133 Roma, Italy}
  \centerline{$^3$Osservatorio Astronomico di Monteporzio}
  \centerline{\elevenrm Via Osservatorio 2, 00040 Roma, Italy}
  \centerline{$^4$Dipartimento di Fisica, Univerit\'a degli studi ``Roma Tre''}
  \centerline{\elevenrm Via della Vasca Navale 84, Roma, I00146 Italy}
\vspace{3cm}
\begin{abstract}

We present photoionization models accounting for both photoelectric
and resonant absorption.  We demonstrate that resonance absorption
lines are detectable even in moderate resolution X-ray spectra of type
1 AGN.  The spectra transmitted by gas illuminated by the ionizing
continua typical of flat X-ray spectrum, broad optical emission line
type 1 AGN, and steep X-ray spectrum, narrow optical emission line
type 1 AGN (NLSy1) are dramatically different. In particolar, we find
that both the absorption features seen in many Seyfert 1 galaxies
between 0.7 and 1 keV and in several NLSy1 between 1 and 2 keV can be
explained by our model, without requiring relativistic outflowing
velocities of the gas.

\end{abstract}
\vspace{2.0cm}

\section{ Introduction }

Soft X-ray spectra of X-ray flat, broad optical emission line type 1
AGN frequently show signatures of reprocessing by ionized matter along
the line of sight (Reynolds, 1997).  The main features imprinted on
these spectra are OVII-OVIII K edges at 0.74 and 0.87 keV.  However,
as recently pointed out by several authors, photoelectric absorption
is not able to entirely account for the complexity of the spectra
observed by even moderate resolution instruments.  Significant
negative residuals between 0.9 and 1 keV, and, in some cases, positive
residuals between 0.5 and 0.7 keV, are still present in ASCA-SIS and
BeppoSAX LECS spectra of several objects (e.g. NGC~985, Nicastro et
al., 1997, 1998a; NGC~3783, George et al., 1998, Piro et al., 1998;
MCG-6-30-15, NGC~3516, Reynolds, 1997, Nicastro et al, 1998b).

On the other hand, deep OVII-OVIII K edges have never been observed in
the spectra of X-ray steep, narrow optical emission line type 1 AGNs
(NLSy1).  Nevertheless, the X-ray spectra of several NLSy1 are not
featureless, showing broad negative residuals between 1 and 2 keV,
after proper modelling of the intrinsic continuum (usually a curved
continuum, flattening above 2-3 keV by $\Delta\alpha\sim0.5-1$, Boller
et al., 1996, Laor et al., 1997, Brandt et al., 1997).
\footnote{IRAS~13224-3809, PG~1404+226, 1H~0707-495, Leighly et al, 1997; 
PG~1244+026, Fiore et al., 1998, Ark~564, Comastri et al., 1997.}
Leighly et al. (1997) interpret these features as blueshifted (by $\sim 
0.5$c) OVII-OVIII resonance absorption lines. 
Fiore et al. (1998) model a similar feature 
in the ASCA spectrum of PG~1244+026 in terms of either an emission line 
at 0.92 keV or Fe XVIII and Fe XVII resonant absorption blueshifted by 
$\sim 0.3$ c.

Here we present photoionization models accounting for both
photoelectric and resonant absorption, and compare them to 
the spectra observed in a broad line Seyfert 1 galaxy and a 
NLSy1 (see Nicastro et al., 1998a for more details). 
We show simulations of our models with the moderate 
resolution ASCA--SIS CCDs, the high spectral resolution gratings 
AXAF--HETG, and the high collecting area, high spectral resolution 
baseline Constellation-X calorimeter. We briefly discuss the relevant 
physics which could be addressed in this field by these instruments. 


\section{Photoelectric+Resonant Absorption Model}
We built single-zone photoionization models accounting for both 
photoelectric and resonant absorption. 
We consider K$\alpha$ and K$\beta$ resonance absorption lines from 
H-like and He-like ions of C, O, Ne, Mg, Si, S and Fe, as well as 
a complex of almost 100 L lines from Si and Fe at $\sim 0.3$ and $\sim 1$ 
keV respectively (atomic data are from Verner et al. 1996). 

We use CLOUDY (Ferland, 1996, vs. 90.01) to calculate the physical structure 
of a given cloud of gas photoionized by a given ionizing continuum shape. 
Intensities, equivalent widths and profiles of the considered resonance 
absorption lines are then consistently computed considering both 
Doppler and natural broadening mechanisms. Doppler broadening includes 
a term accounting for turbulence of the gas along the line of sight: 
\begin{equation} \label{dopplerwidth}
\Delta\nu_D = {\nu_0 \over c} \left( {{2kT} \over M} + \sigma_v^2 
\right)^{0.5}. 
\end{equation}
In order to evaluate the relative contribution of these two terms 
to the total optical depth, we report the ratio between the central 
optical depth in the two extreme cases: $\sigma_v=0$ (thermal motion 
only: $\tau_0^{therm}$) and $\sigma_v \gg \sqrt{kT/M}$. 
This quantity does not depend on the particular considered transition, 
the distribution of the ionic abundances in the gas, and its column 
density, and is therefore a good estimator of the relative contribution 
of the two Doppler broadening mechanisms to the core optical depth: 
\begin{equation} \label{tau0thermtau0turb}
{{\tau_0^{therm}} \over {\tau_0^{turb}}} = 2.3 A^{0.5} (T)_6^{-0.5} 
(\beta_{\sigma_v})_{-3}. 
\end{equation}
In eq. (\ref{tau0thermtau0turb}) A is the atomic weight of the given 
element, and $\beta_{\sigma_v} = \sigma_v / c$. 
A given column of non-turbulent highly ionized gas in photoionization 
equilibrium can therefore be more than a factor 10 thicker to the 
resonant absorption process (at the energy of the transition) 
than an identical column of gas undergoing strong turbulence ($\sigma_v 
\ge 300$ km s$^{-1}$, i.e. $(\beta_{\sigma_v})_{-3} \ge 1$). 

\subsection {Transmitted Spectra}

As an example we present two spectra transmitted by gas photoionized
by two drastically different ionizing continua: a flat X-ray continuum
with $\alpha_E=0.9$, and a double component X-ray continuum including
a low energy component, parameterized by a black-body with kT=0.15
keV, and a power law component with $\alpha_E=0.9$. Similar continua
are typically observed in broad line type 1 AGN and NLSy1
respectively.  Fig. 1a and 1b show these spectra. The 2-10 keV flux of
these spectra is of $10^{-11}$ erg s$^{-1}$ cm$^{-2}$.  In both cases
we include a cut-off due to neutral absorption from a column of
$N_H=3\times 10^{20}$ (to simulate Galactic absorption). The
dispersion and outflowing gas velocities are $\sigma_v = 500$ km
s$^{-1}$, and v=1000 km s$^{-1}$ respectively, typical of the UV/X-ray
ionized absorbers (Mathur et al., 1994, 1995, 1998).

The spectrum in Fig. 1a (Ionization Parameter U=1, N$_H=10^{22}$ cm$^{-1}$) 
shows a complex absorption structure between 0.3 and 2 keV whose strongest 
features are CVI, OVII-OVIII and NeIX-NeX K-edges, at 0.49 keV, 
0.74-0.87 keV and 1.20-1.36 keV respectively. 
These features have been observed in moderate resolution spectra of 
almost half of the flat X-ray spectrum, broad optical emission lines 
Seyfert 1 galaxies observed by ASCA (Reynolds et al., 1997). 
However other relevant features are clearly visible in this spectrum. 
In particular we note the presence of strong K$\alpha$--K$\beta$ 
resonance absorption lines from H-like and He-like ions of O and Ne, and 
two systems of L resonance absorption lines from Si ($\sim 0.3-0.4$ keV) and 
Fe ($\sim 1-2$ keV). These lines could well be present in the spectra of 
many  known warm absorber AGNs.
The energy of several of the Fe-L and Ne lines is close 
to that of the OVIII K-edge, and their integrated equivalent width 
can be as high as  $\sim 20$ eV for turbolence velocities 
of $\sim 500$ km s$^{-1}$. Not including them in the modeling
of even moderate resolution spectra could lead to a significant
overestimate of the OVIII K-edge optical depth.

\bigskip
The spectrum in Fig. 1b (U=10, N$_H=10^{23}$ cm$^{-1}$) does not show 
any absorption K-edges in the 0.1-3 keV range, despite of the large column 
density used. This is a direct consequence of the much steeper soft X-ray 
ionizing continuum (compared to the one used to produce the spectrum 
in Fig. 1a). 
This continuum, in fact, contains enough photons to fully ionize all the 
elements lighter than Ne-Mg. Iron ions instead are still distributed 
in medium-high ionization states. 
This gives rise to a complex system of resonant absorption 
lines and edges from the Iron L shells, all between 1 and 2 keV.
These absorption features may therefore be responsible for the
deficit of counts seen by ASCA in several NLSy1 at these energies, 
If this is the case, the gas outflowing velocity would not need
to be relativistic.
The ionization parameter adopted (U = 10) is expected for sources of 
the luminosity of the NLSy1 of the Leighly et al. (1997) sample 
provided that: (a) the physical properties of the absorber are similar 
to those of the absorbers usually seen in broad line type 1 AGN; and 
(b) the radial distance of the gas from the ionizing sources, in units 
of gravitational radii, scales approximatively with the square root of 
the luminosity (as in fact observed  in the Reynolds, 1997,  sample). 
\begin{figure}
\epsfxsize=13cm 
\hspace{3.5cm}\epsfbox{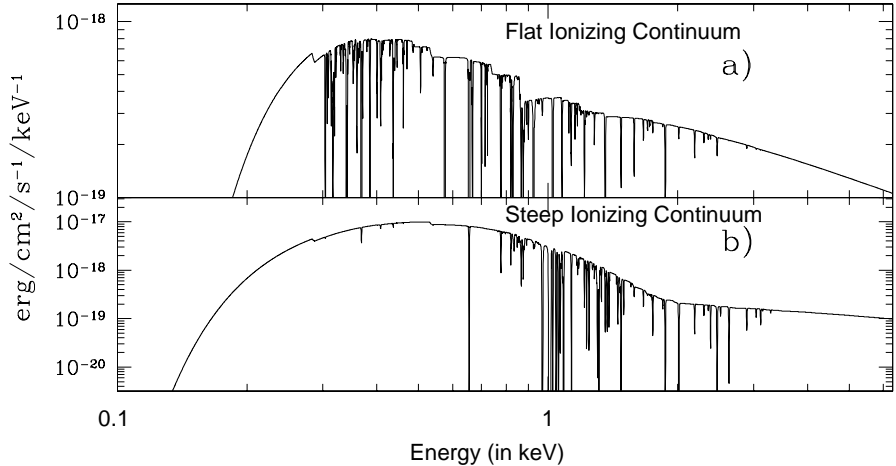} 
\vspace{-6truecm}
\caption[h]{Spectra emerging from clouds of gas photoionized by (a) 
a typical Seyfert 1 continuum, and (b) a typical NLS1 continuum (see 
details in the text).}
\end{figure}

\subsection{Simulated Spectra}

For both the transmitted spectra of Fig. 1a and 1b we carried out
100 ksec ASCA-SIS
\footnote{$\Delta E = 100$ eV, collecting area $\sim 100$ cm$^2$, at 
1 keV, (``The ASCA Data Reduction Guide'', vs. 2.0, 1996).}
, 100 ksec AXAF-HETG 
\footnote{$\Delta E = 1.5$ eV, collecting area $\sim 100$ cm$^2$, at 
1 keV, (``AXAF Proposer's Guide'', vs 1.0, 1997).}
and 20 ksec Constellation-X calorimeter 
\footnote{$\Delta E = 3$ eV, collecting area $\sim 10,000$ cm$^2$, 
(``The High Throughput X-ray Spectroscopy (HTXS) Mission'', 1997).} 
simulations. 
We performed the AXAF-HETG simulations with the MARX simulator (``MARX User 
Guide'', vs 1.0, 1997), while, to simulate the ASCA-SIS and the 
Constellation-X calorimeter spectra, we wrote a simple program to fold 
the emerging spectra of Fig. 1a and 1b through the instrument responses
\footnote{ftp://legacy.gsfc.nasa.gov/caldb/data/asca/sis/bcf/; 
ftp://legacy.gsfc.nasa.gov/htxs/calor.rsp}
, and add background (for the ASCA-SIS) and statistical noise. 

In the upper panels of Fig. 2 and 3 we show the ratios between the 
two ASCA-SIS simulated spectra and the continuum used in the
simulations (i.e. without the resonance absorption lines).
Clear negative residuals at 0.6-1 keV and 1-2 keV respectively are 
present in both the panels, so indicating the capability of even moderate 
resolution instruments to detect resonance absorption lines. 

Fig. 3bc and 4bc show the AXAF-HETG and Constellation-X simulations 
of the models in Fig 1a and 1b respectively. 
The high resolution AXAF-HETG resolves most of the lines and edges 
present in the emerging spectra of Fig. 1a and 1b, so allowing a stringent 
test of our ``ionized absorber+resonant absorption'' models. 
Despite of the low signal to noise of the first order HETG spectra, 
outflowing or inflowing velocity of the gas can be accurately determined, 
as well as turbolence velocity along the line of sight. 
The much higher (factor of 100) effective area of the baseline 
Constellation-X calorimeter, will permit to accumulate very good 
quality spectra of these (and at least one order of magnitude 
lower luminosity) AGNs with much shorter (factor of 5) exposure time, 
so allowing for: a) detailed variability studies of the ionization 
state of the gas; b) an accurate measure of absorption line ratios,  which
in turn would allow the use of powerful plasma diagnostic; c) the high 
resolution observation of a large sample of AGN, which would be useful 
for the determination of the distribution of the absorber density, 
ionization state, chemical composition and distance from the ionizing 
source and for the study of the evolution of these distributions with
redshift and luminosity.  
\begin{figure}
\epsfxsize=9cm 
\hspace{2.5cm}\epsfbox{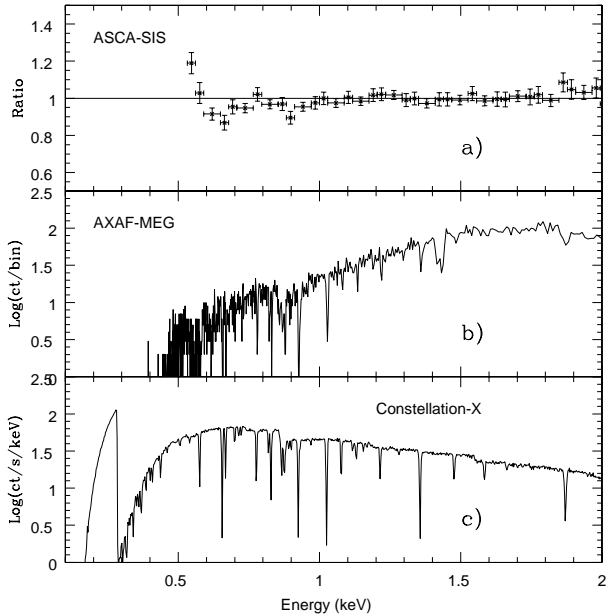} 
\caption[h]{Ratio between a 100 ksec ASCA-SIS simulation of the 
model in Fig. 1a and the best fit warm absorber model accounting 
for photoelectric absorption only (a). 100 ksec AXAF-MEG and 20 
ksec Constellation-X simulations of the emerging spectrum in Fig. 1a 
(b and c respectively).}
\end{figure}
\begin{figure}
\epsfxsize=9cm 
\hspace{2.5cm}\epsfbox{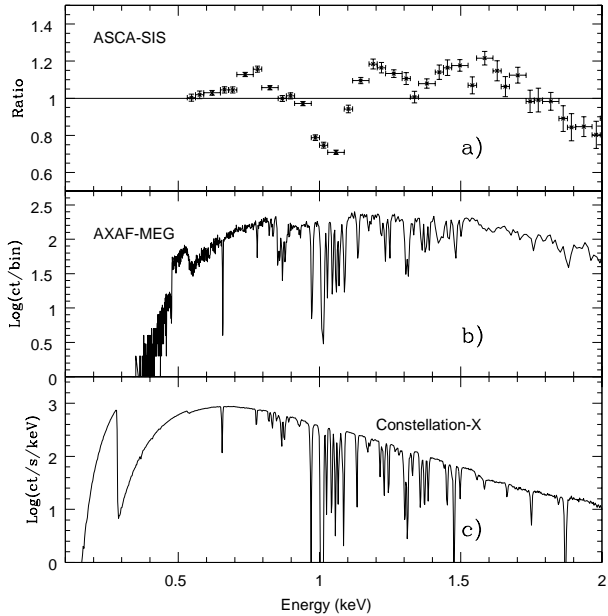} 
\caption[h]{Ratio between a 100 ksec ASCA-SIS simulation of the 
model in Fig. 1b and the best fit continuum model (a). 
100 ksec AXAF-MEG and 20 ksec Constellation-X simulations of the 
emerging spectrum in Fig. 1b (b and c respectively).}
\end{figure}

\section {Conclusion}

We have calculated spectra emerging from ionized gas illuminated 
by flat and steep ionizing continua, typical of Seyfert galaxies and
NLSy1 respectively, taking into account both photoelectric and resonant 
absorption.
We have convolved these spectra with the responses of the moderate resoltion 
ASCA-SIS CCDs, the high resolution high energy trasmittion gratings on board 
AXAF (AXAF-HETG), and the  high collecting area, good resolution 
baseline Constellation-X calorimeter.  Our major findings are: 

\begin{enumerate}
\item The distribution of relative ionic abundances in photoionized 
gas dramatically depends on the exact shape of the ionizing continuum. 

\item Strong K$\alpha$ and K$\beta$ resonance absorption lines from He-like 
and H-like ions of C, O, and Ne, along with deep CVI, OVII-OVIII, NeIX-X K 
edges, and FeXV-XVII L edges, are expected to be present in the 0.1-3 keV 
spectra of flat X-ray spectrum quasars emerging from photoionized gas along 
the line of sight. When fitting low and medium resolution X-ray data with 
phenomenological multi-edge models, these resonance absorption lines may 
produce the effect of overestimating the optical depth of the OVIII K edge. 
 
\item Conversely, no strong absorption edge is expected to be imprinted 
on spectra emerging from gas illuminated by steep ionizing continua and 
with physical and geometrical properties similar to those found in broad 
line type 1 AGN. However, a large number of strong Fe L and Mg, Si, and 
S K$\alpha$ and K$\beta$ resonance absorption lines (almost 70) are 
predicted between 1 and 2 keV. 
Fitting low and medium energy resolution data with continuum models may 
then result in deep smooth negative residuals at these energies.
We then suggest that the $\sim 1$ keV absorption feature observed 
by ASCA in several Narrow Line Seyfert 1 galaxies can be explained
in terms of resonance absorption lines, without requiring relativistic 
outflowing velocities of the gas.

\item Future observations with the high resolution gratings on 
board AXAF will allow to unanbiguously detect the most important 
resonance absorption lines and then measure the gas dispersion velocity.

\item The observations possible with the Constellation calorimeter 
will allow the application of powerful plasma diagnostics to the
brightest AGN and the study of a large sample of AGN for statistical 
purposes.

\end{enumerate}

\section { References}

\refe Boller, Th., Brandt, W.N \& Hink, H,: 1996, \AAP 305 53. 

\refe Brandt, W.N., Mathur, S., Elvis, M.: 1997, \MN 285 25. 

\refe Comastri, A., et al.: 1997, Proceedings of the Symposium 
``The active X-ray sky'', {\bf in press, astro-ph/9712278.} 

\refe George, I.M., Turner, T.J., Mushotzky, R., Nandra, K., Netzer, H.: 
1998, \APJ 503 174. 

\refe Ferland G.J., 1996 CLOUDY vs. 90.01

\refe Fiore, F., et al.: 1998, \MN 298 103. 

\refe Laor, A., Fiore, F., Elvis, E., Wilkes, B.J., McDowell, J.C.: 
1997, \APJ 477 93. 

\refe Leighly, K.M., Mushotzky, R.F., Nandra, K. \& Forster, K.: 
1997, \APJL 489 25. 

\refe Mathur S., Wilkes B.J., Elvis M., Fiore F.: 1994, \APJ 434 493

\refe Mathur S., Elvis M., Wilkes B.J.: 1995, \APJ 452 230

\refe Mathur S., Wilkes B.J., Aldcroft T.: 1998, {\it Astrophys. J} 
{\bf in press.}

\refe Nicastro, F., Fiore, F., Brandt, W.N., Reynolds, C.S.: 1997, 
Proceedings of the Symposium ``The active X-ray sky'', {\bf in press.} 

\refe Nicastro, F., Fiore, F. \& Matt G.: 1998a, {\it Atrophys. J.} 
{\bf submitted.}

\refe Nicastro, F. \& Fiore, F.: 1998b, {\bf in preparation.} 

\refe Piro, L., Nicastro, F., et al.: 1998, {\bf in preparation.} 

\refe Reynolds, C.S.: 1997, \MN 287 513

\refe Verner, D. A., Verner, E. M., and Ferland, G. J., 1996, 
{\it Atomic Data Nucl. Data Tables}

\end{document}